\documentclass[a4paper,fleqn]{cas-dc}
\usepackage[numbers]{natbib}
\usepackage{amsmath,amssymb}
\usepackage[cp1251]{inputenc}

\usepackage{srctex}
\usepackage{hyperref}

\begin{document}
\let\WriteBookmarks\relax
\def\floatpagepagefraction{1}
\def\textpagefraction{.001}
\shorttitle{Al-Cu-Fe alloys: the relationship between the quasicrystal and its melt}
\shortauthors{L.V.~Kamaeva et~al.}

\title [mode = title]{Al-Cu-Fe alloys: the relationship between the quasicrystal and its melt}
\tnotemark[1]

\tnotetext[1]{This document is the results of the research
   project funded by the Russian Science Foundation.}

\author[1,2]{L.V.~Kamaeva}
\cormark[1]
\ead{lara_kam@mail.ru}

\author[3,4,2]{R.E.~Ryltsev}

\author[1]{V.I.~Lad`yanov}

\author[2,4,5]
{N.M.~Chtchelkatchev}[orcid=0000-0002-7242-1483]
\cormark[2]
\ead[url]{shchelkachev.nm@mipt.ru}

\address [1]{Udmurt Federal Research Center, Ural Branch of Russian Academy of Sciences, 426068 Izhevsk, Russia}
\address [2]{Vereshchagin Institute for High Pressure Physics, Russian Academy of Sciences, 108840 Troitsk, Moscow, Russia}
\address [3]{Institute of Metallurgy, Ural Branch of Russian Academy of Sciences, 620016, Ekaterinburg, Russia}
\address [4]{Ural Federal University, 620002, Ekaterinburg, Russia}
\address [5]{Moscow Institute of Physics and Technology, 141700, Dolgoprudny, Moscow Region, Russia}

\cortext[cor1]{Principal corresponding author}

\begin{abstract}
Understanding the mechanisms which relate properties of liquid and solid phases is crucial for fabricating new advanced solid materials, such as glasses, quasicrystals and high-entropy alloys.  Here we address this issue for quasicrystal-forming Al-Cu-Fe alloys which can serve as a model for studying microscopic mechanisms of quasicrystal formation. We study experimentally two structural-sensitive properties of the liquid -- viscosity and undercoolability -- and compare results with \textit{ab initio} investigations of short-range order (SRO). We observe that SRO in Al-Cu-Fe melts is polytetrahedral and mainly presented by distorted Kasper polyhedra. However, topologically perfect icosahedra are almost absent an even stoichiometry of icosahedral quasicrystal phase that suggests the topological structure of local polyhedra does not survive upon melting. It is shown that the main features of interatomic interaction in Al-Cu-Fe system, extracted from radial distribution function and bong-angle distribution function, are the same for both liquid and solid states. In particular,  the system demonstrates pronounced repulsion between Fe and Cu as well as strong chemical interaction between Fe and Al, which are almost concentration-independent. We argue that SRO and structural-sensitive properties of a melt may serve as useful indicators of solid phase formation.  In particular, in the concentration region corresponding to the composition of the icosahedral phase, a change in the chemical short-range order is observed, which leads to minima on the viscosity and udercoolability isotherms and has a noticeable effect on the initial stage of solidification.
\end{abstract}

\begin{graphicalabstract}
\centering
\includegraphics{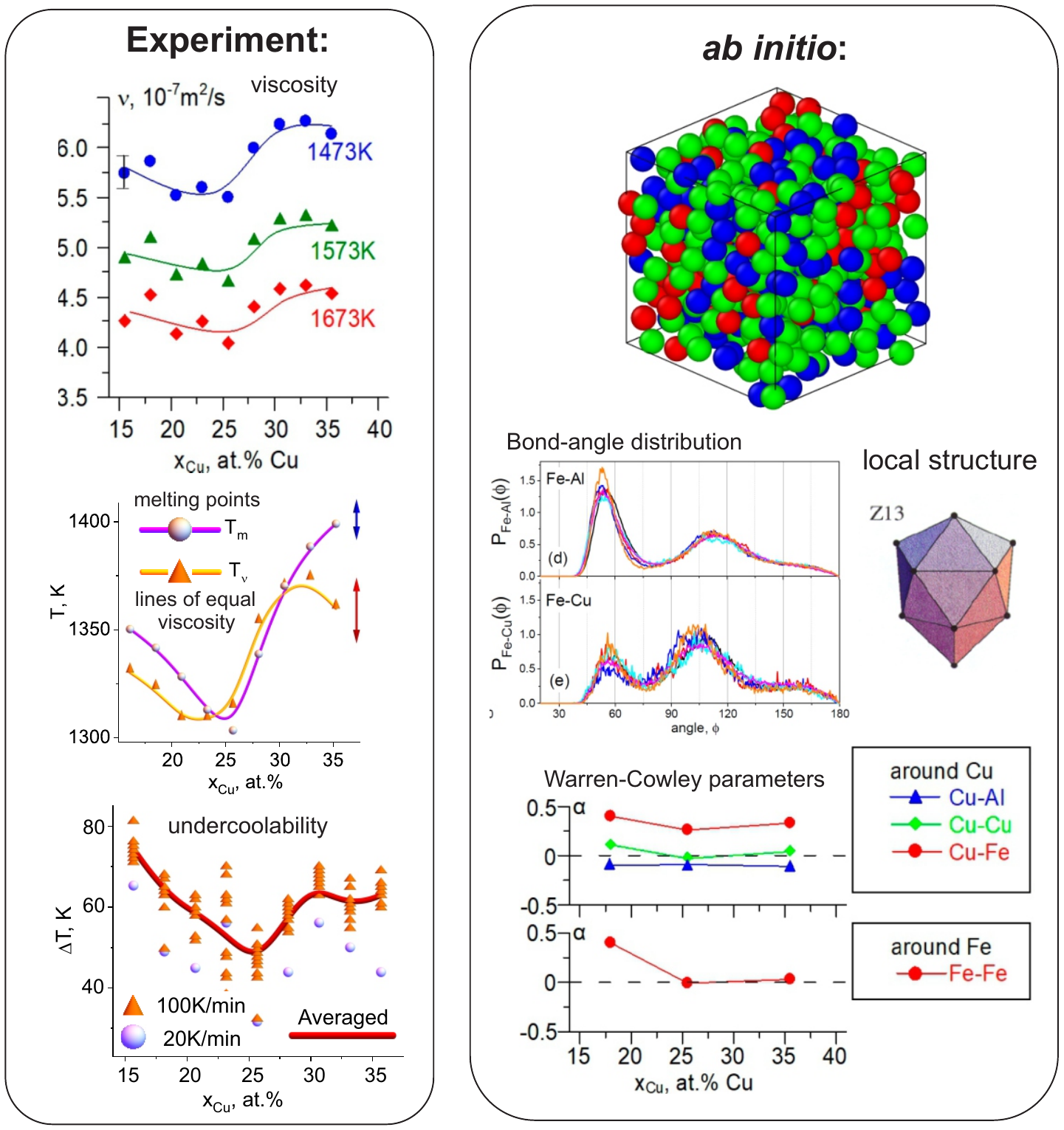}
\end{graphicalabstract}

\begin{highlights}

\item We find relation between structure of Al-Cu-Fe melts, their structural sensitive properties and tendency to form ico-phase
\item Isotherms of viscosity and undercoolability of Al-Cu-Fe melts develop minima at ico-phase stoichiometry
\item Short-range order in Al-Cu-Fe melts is polytetrahedral and presented by distorted Kasper polyhedra
\item Local icosahedra in Al-Cu-Fe quasicrystal phase are destroyed by melting but pair and angular correlations survive
\item Chemical short-range order of Al-Cu-Fe melts changes in the vicinity of i-phase stoichiometry

\end{highlights}


\maketitle

\section{Introduction}

Structural heredity relating properties of liquid and solid phases is one of the topical issues in modern material science~\cite{Bendert2012PRL, Ryltsev2015SoftMatt,Wu2016SciRep,Ryltsev2017SoftMatt}. For example, the idea that the structure of a melt can be related to the tendency to form certain solid phases was recently applied to high-entropy alloys~\cite{Gao2013Entropy,Santodonato2015NatureComm,Ding2018ApplPhysLett}. Recently, a method for predicting the formation of layered quasicrystals from structural characteristics of fluid phase has been proposed~\cite{Ryltsev2015SoftMatt,Ryltsev2017SoftMatt}. The relationship between structural-sensitive liquid phase properties (viscosity, thermal expansion etc.) and glass-forming ability in metallic alloys has been intensively discussed~\cite{Bendert2012PRL,Godbole2004JAC,Terzieff2008JAC,Bendert2013JNonCrystSol,Yakimovuch2014JAC,Pan2015JAC,Kelton2016JPCM,Johnson2017JAC,Dubinin2019JAC,Filippov2019JAC,Shi2019JAC}.

It is known that metallic alloys, especially Al-based ones, can demonstrate complex structure in liquid state~\cite{Muratov2014JMolLiq,Xie2019NatureComm,Wang2018JMolLiq,Roik2010JMolLiq,wang2017115,Debela2018173}. In particular, it has been suggested that the unusual structure of Al-based melts leads to the formation of complex crystalline and quasicrystalline phases~\cite{wang2017115,Debela2018173}. That is especially relevant for the Al-Cu-Fe system which can form stable icosahedral phase (i-phase) as well as complex crystal phases~\cite{Holland1997976}.



Al-Cu-Fe quasicrystalline alloys are one the most promising  quasicrystal metal materials for practical use~\cite{Huttunen-Saarivirta2004JALCOM}, therefore the stable i-phase in this system is rather well studied~\cite{Inoue1989MaterTransJIM,LEE2001871,Holland-Moritz1999829,Faudot1991383,WOLF20181288,leskovar2018epitaxial,kang2018situ,gharehbaghi2018experimental,kawazoe2019structural,wang2019synthesis,tcherdyntsev2019formation,wang2019synthesis}. The Al-Cu-Fe i-phase can be obtained by rapid quenching of a melt (for example, by spinning or sputtering technique), mechanical alloying and plasma deposition methods~~\cite{WOLF20181288,Salimon2017315,Barua2001863}. These methods produce i-phase in the form of either powders or thin coatings those applicability is essentially restricted. Therefore, the search for methods to fabricate bulk single-phase quasicrystalline Al-Cu-Fe alloys is currently an important task. Good results in this direction can be achieved by using different costly technologies, for example, the preparation of powders and their subsequent high-tech heat treatment~\cite{Nicula2008113,Srivastava2014258}. Classical metallurgical casting technologies do not allow to obtain a single-phase material with controlled composition and property~\cite{Tsai2008,LEE2001871,Yokoyama200068}, although there are some successes in this area~\cite{Yokoyama200068,dolinvsek2007PRB}. A promising way to solve these problems is the use of structural heredity between liquid and solid states.

The structure of Al-Cu-Fe melts is satisfactorily investigated in only the vicinity of i-phase stoichiometry \cite{Holland1997976,qin2011quasicrystal,Xue-Lei700}. The data available reveal pronounced icosahedral short-range order (SRO) in this composition range. However, the existence of such SRO at other compositions as well as its impact on the solidification process are still almost unexplored.

Viscosity is one of the main structural-sensitive properties of a fluid. For metal melts, it is also of practical importance because the viscosity value affects significantly the casting process.  Moreover, the analysis of the viscosity allows estimating qualitatively how the interatomic interaction in a melt varies with a change of temperature or atomic concentration~\cite{Sterkhova2014241,Beltyukov20151,Sterkhova2014250,Kamaeva2012InorgMater}. Another useful structural-sensitive property is undercoolability $\Delta T$  whose concentration dependencies allow understanding the solidification process in multicomponent alloys. On the one hand, $\Delta T$ depends on the initial structure of a liquid phase, and on the other hand, it controls the concurrent processes of nucleation and growth of a solid phase.

Previously Al-Cu-Fe melts were studied in the vicinity of particular compositions. The viscosity of Al-Cu-Fe melts was investigated for the composition ${\rm Al_{63}Cu_{25}Fe_{12}}$ in the temperature range from 1100 to 1300$^\circ$C~\cite{qin2011quasicrystal}; the effect of supercooling on the phase formation sequence was studied for ${\rm Al_{62}Cu_{25.5}Fe_{12.5}}$  and ${\rm Al_{60}Cu_{34}Fe_{6}}$ alloys in~\cite{Holland1997976}.

Here we study Al-Cu-Fe melts in a wide range of compositions. We consider the following two concentration cross-sections: Al$_{57.0+x}$Cu$_{30.5-x}$Fe$_{12.5}$ and Al$_{52.0+x}$Cu$_{25.5}$Fe$_{22.5-x}$, (where $x=0-20$ at.\%), containing i-phase stoichiometry composition. For these alloys, we study experimentally two structural-sensitive properties of the liquid -- viscosity and undercoolability -- and compare results with \textit{ab initio} investigations of SRO.

\section{Methods}

\begin{table*}[pos=h]
\centering
\caption{Parameters of Eq.~\eqref{GrindEQ1} for temperature dependencies $\nu$ of Al-Cu-Fe melts.}
\label{tab:table1}
\begin{tabular}{|l|c|c||l|c|c|}
\hline
composition       & A$_v$, 10$^{-8}$ m$^2$/s       & E$_v$, kJ/mol & composition     & A$_v$, 10$^{-8}$ m$^2$/s       & E$_v$, kJ/mol \\ \hline
${\rm Al_{72}Cu_{25.5}Fe_{2.5}}$   & 7.3                            & 18.8 &  ${\rm Al_{67}Cu_{20.5}Fe_{12.5}}$       & 4.9                            & 29.6       \\ \hline
${\rm Al_{67}Cu_{30}Fe_{2.5}}$     & 5.9                            & 21.8  &  ${\rm Al_{64.5}Cu_{23}Fe_{12.5}}$      & 5.7                            & 28.0       \\ \hline
${\rm Al_{62}Cu_{35.5}Fe_{2.5}}$   & 6.4                            & 21.4  &  ${\rm Al_{62}Cu_{25.5}Fe_{12.5}}$      & 4.2                            & 31.5       \\ \hline
${\rm Al_{60.5}Cu_{35.5}Fe_{4}}$   & 5.2                            & 24.7   & ${\rm Al_{59.5}Cu_{28}Fe_{12.5}}$      & 4.6                            & 31.5       \\ \hline
${\rm Al_{79.7}Cu_{14.5}Fe_{5.8}}$  & 8.4                            & 21.1 &  ${\rm Al_{57}Cu_{30.5}Fe_{12.5}}$      & 4.8                            & 31.3       \\ \hline
${\rm Al_{73.7}Cu_{20.5}Fe_{5.8}}$  & 6.7                            & 22.1  &  ${\rm Al_{54.5}Cu_{33}Fe_{12.5}}$     & 4.9                            & 31.1       \\ \hline
${\rm Al_{68.7}Cu_{25.5}Fe_{5.8}}$  & 5.2                            & 24.9 &  ${\rm Al_{52}Cu_{35.5}Fe_{12.5}}$       & 5.0                            & 30.8       \\ \hline
${\rm Al_{63.7}Cu_{30.5}Fe_{5.8}}$  & 6.0                            & 23.3  &  ${\rm Al_{70.6}Cu_{15.7}Fe_{13.7}}$    & 4.2                            & 33.2       \\ \hline
${\rm Al_{58.7}Cu_{35.5}Fe_{5.8}}$  & 5.1                            & 26.3 &   ${\rm Al_{68.1}Cu_{18.2}Fe_{13.7}}$    & 5.0                            & 30.8       \\ \hline
${\rm Al_{53.7}Cu_{40.5}Fe_{5.8}}$  & 5.4                            & 26.1 &  ${\rm Al_{60.8}Cu_{25.5}Fe_{13.7}}$     & 4.2                            & 31.7       \\ \hline
${\rm Al_{67}Cu_{25.5}Fe_{7.5}}$   & 4.3                            & 25.7  &  ${\rm Al_{59.5}Cu_{25.5}Fe_{15}}$      & 4.0                             & 32.9       \\ \hline
${\rm Al_{64.5}Cu_{25.5}Fe_{10}}$   & 3.9                            & 30.3 &   ${\rm Al_{57}Cu_{25.5}Fe_{17.5}}$     & 4.6                            & 33.2       \\ \hline
${\rm Al_{72}Cu_{15.5}Fe_{12.5}}$  & 4.8                            & 30.4  &   ${\rm Al_{54.5}Cu_{25.5}Fe_{20}}$     & 3.9                            & 36.2       \\ \hline
${\rm Al_{69.5}Cu_{18}Fe_{12.5}}$   & 6.7                            & 26.4  & ${\rm Al_{52}Cu_{25.5}Fe_{22.5}}$      & 4.5                            & 34.4       \\ \hline
\end{tabular}
\end{table*}


Samples were obtained by fusing of Al-Fe master alloy (either ${\rm Al_{80.6}Fe_{19.4}}$ or ${\rm Al_{69.8}Fe_{30.2}}$ depending on the melted composition), electrolytic aluminum and cathode copper in a viscosimeter furnace in an inert atmosphere of He (after preliminary pumping to 10-2Pa) at temperature 1400$^\circ$C for 1~hour.

Kinematic viscosity $\nu$ was measured by the method of damped torsional vibrations of a corundum crucible with a melt~\cite{Terzieff2008JAC}. The method is described in detail in~\cite{Beltyukov2008}. The experiments were carried out in an atmosphere of purified helium by using crucibles of aluminum oxide. To prevent an uncontrollable influence of the oxide surface film, a cover of ${\rm Al_2O_3}$ was placed on the melt surface and fixed so that it served as a second end surface of the crucible~\cite{Beltyukov2015}. Preliminary experiments have shown that, immediately after melting, alloys under consideration demonstrate slow relaxation of the viscosity (2 hours or more). The reason is the slow establishment of thermodynamic equilibrium in the melt -- crucible -- atmosphere system~\cite{Kamaeva2018}. This feature must be taken into account when measuring the temperature dependencies of viscosity $\nu (T)$. Therefore, measurements of $\nu (T)$ are performed in the following mode. After melting, the liquid is overheated by 50$^\circ$ C above the liquidus temperature, kept at this temperature for 10 minutes and then cooled to the liquidus temperature. Then, we perform viscosity measurements by heating a system in a stepwise manner from the melting temperature up to 1400$^\circ$ C  by increments of 20-30$^\circ$ and then cooling it  down to melting temperature; at each temperature, isothermal exposure of 7 min is performed before measurement. Next, the samples were cooled down to room temperature and described stages were cyclically  repeated. For all investigated melts, in whole temperature ranges, temperature dependencies $\nu (T)$ obtained in both heating and cooling modes, are well fitted by the Arrhenius relation:
\begin{equation} \label{GrindEQ1}
\nu =A_{\nu } {\mathop{e}\nolimits^{\frac{E_{\nu } }{RT} }} ,
\end{equation}
where $A_{\nu }$ is a pre-exponential factor, $E_{\nu }$ is the activation energy, $R$ is the universal gas constant, $T$ is the absolute temperature. To construct concentration dependencies of the viscosity for each alloy, we fit the viscosity data obtained in two subsequent heating/cooling cycles by common exponential dependence. The corresponding parameters of the equation \eqref{GrindEQ1} for the studied alloys are presented in Table~\ref{tab:table1}.

For a confidence probability of 0.95, the most probable error in determining the absolute values of the viscosity of Al-Cu-Fe melts in a single experiment is 2.5\% with a total error of no more than 5\%.

Both the melting point and the undercoolability of the alloys were determined by using a high-temperature thermal analyzer~\cite{Sterkhova2014241}. The measurements were performed in an inert atmosphere of purified helium under low excess pressure after pre-evacuation to a pressure of the order of $ 10^ {- 2} $ Pa.

DTA plots (thermograms) were obtained in heating mode at the rate of 20~K/min to the selected melt temperature and subsequent cooling, after 20~minutes exposure at maximum temperature. The temperatures of all stages of melting (in heating mode) and crystallization (in cooling mode) for each alloy were  determined from DTA thermograms. Undercoolability was determined as the difference between the liquidus temperature and crystallization start temperature determined from the heating  and cooling thermograms, respectively.

In the process of melt crystallization in a container, the undercoolability value is influenced by different factors, such as cooling rate, the temperature of the molten alloys and the number of melting-crystallization cycles \cite{DT}. Therefore, the experimental measurement process for each sample involved a few heating (melting)-cooling (crystallization) cycles in which either the maximum heating temperature of melts ($T_{\rm max}$) or the cooling rate was changed. Experiments were performed at cooling rates of 20 and 100~K/min. The effect of melt temperature on undercoolability was studied during thermal cycling of the samples. In these experiments, each sample was heated to a temperature of 10–20~K above the liquidus temperature kept at this temperature for 20~minutes, then cooled at the rate of 100~K/min. In the next heating - cooling cycle, the maximum melt temperature was increased by 10-20~K, etc. until the latter was reached 1950~K.

\begin{figure}
  \centering
  \includegraphics[width=0.9\columnwidth]{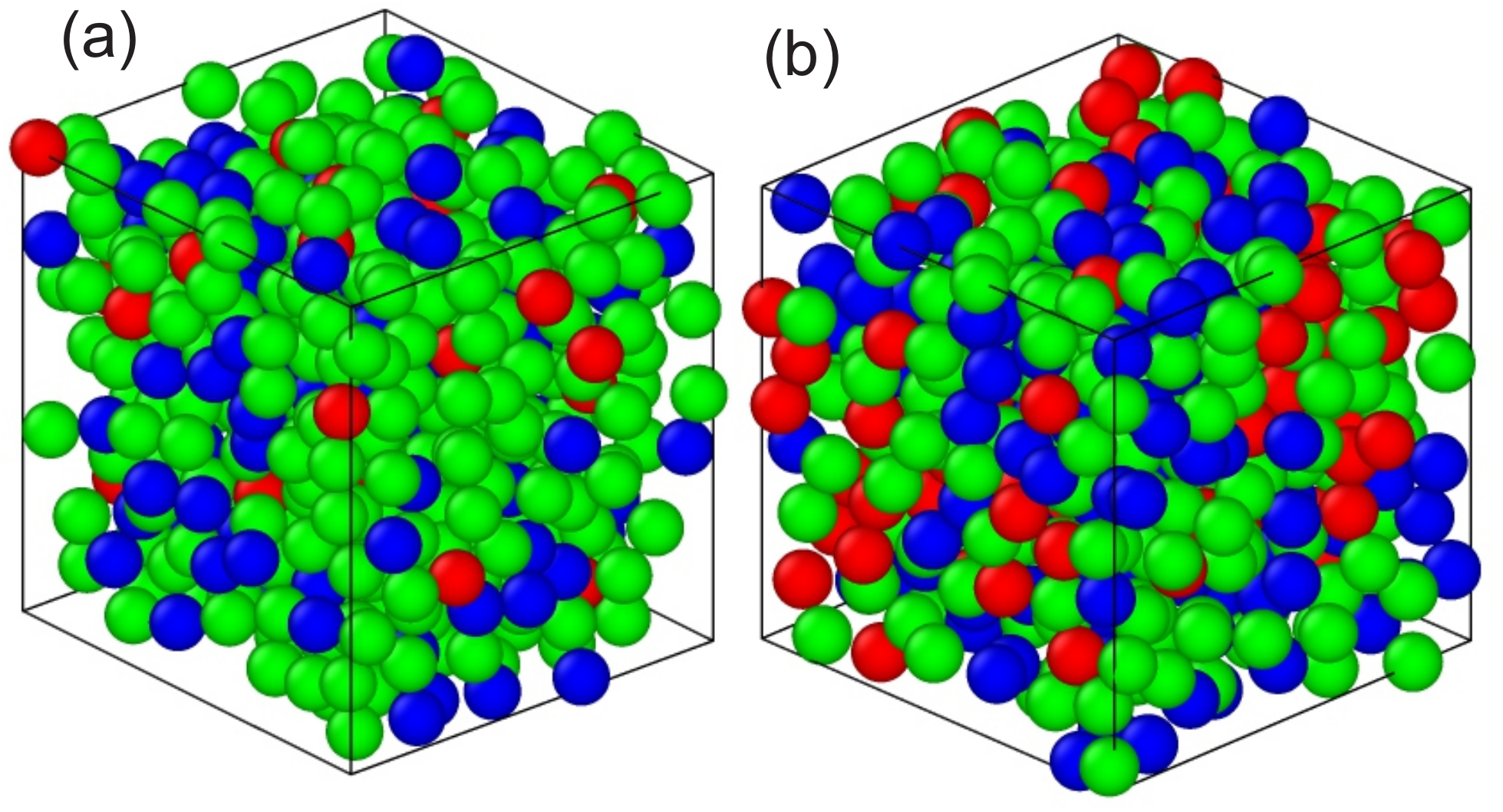}
 \caption{Typical snapshots of atom distribution in ${\rm Al_{68.7}Cu_{25.5}Fe_{5.8}}$ (a) and ${\rm Al_{52}Cu_{25.5}Fe_{22.5}}$ (b) melts obtained by \textit{ab initio} molecular dynamics simulations. Here Al, Cu and Fe are colored green, blue and red, respectively.}\label{fig:particles}
\end{figure}

The structural characteristics of the melts were determined by using \textit{ab initio} molecular dynamics (AIMD) simulations. The calculations were performed by using the open-source CP2K software package \cite{Hutter2014CompMolSci}. Projector augmented-wave (PAW) pseudopotentials and Perdew-Burke-Ernzerhof (PBE) gradient approximation to the exchange-correlation functional were used \cite{Kresse1999PRB}. Calculations were performed in the NVT ensemble at the density corresponding to zero pressure $P = 0$, which was estimated by minimizing energy as a function of volume. The time step was 1~fs. Cubic supercells of 512 atoms (see Fig.~\ref{fig:particles}) with the following compositions were built: Al$_{68.7}$Cu$_{25.5}$Fe$_{5.8}$, Al$_{69.5}$Cu$_{18}$Fe${}_{12.5}$,  Al$_{62}$Cu$_{25.5}$Fe$_{12.5}$,  Al$_{52}$Cu$_{35.5}$Fe$_{12.5}$,  Al$_{57}$Cu$_{25.5}$Fe$_{17.5}$,  and Al$_{52}$Cu$_{25.5}$Fe$_{22.5}$.

These initial configurations were heated to 5000~K and relaxed during several thousand MD steps, then slowly cooled to the target temperatures and relaxed for several thousand MD steps.

\section{Results}

\begin{figure}[pos=t]
  \centering
  \includegraphics[width=\columnwidth]{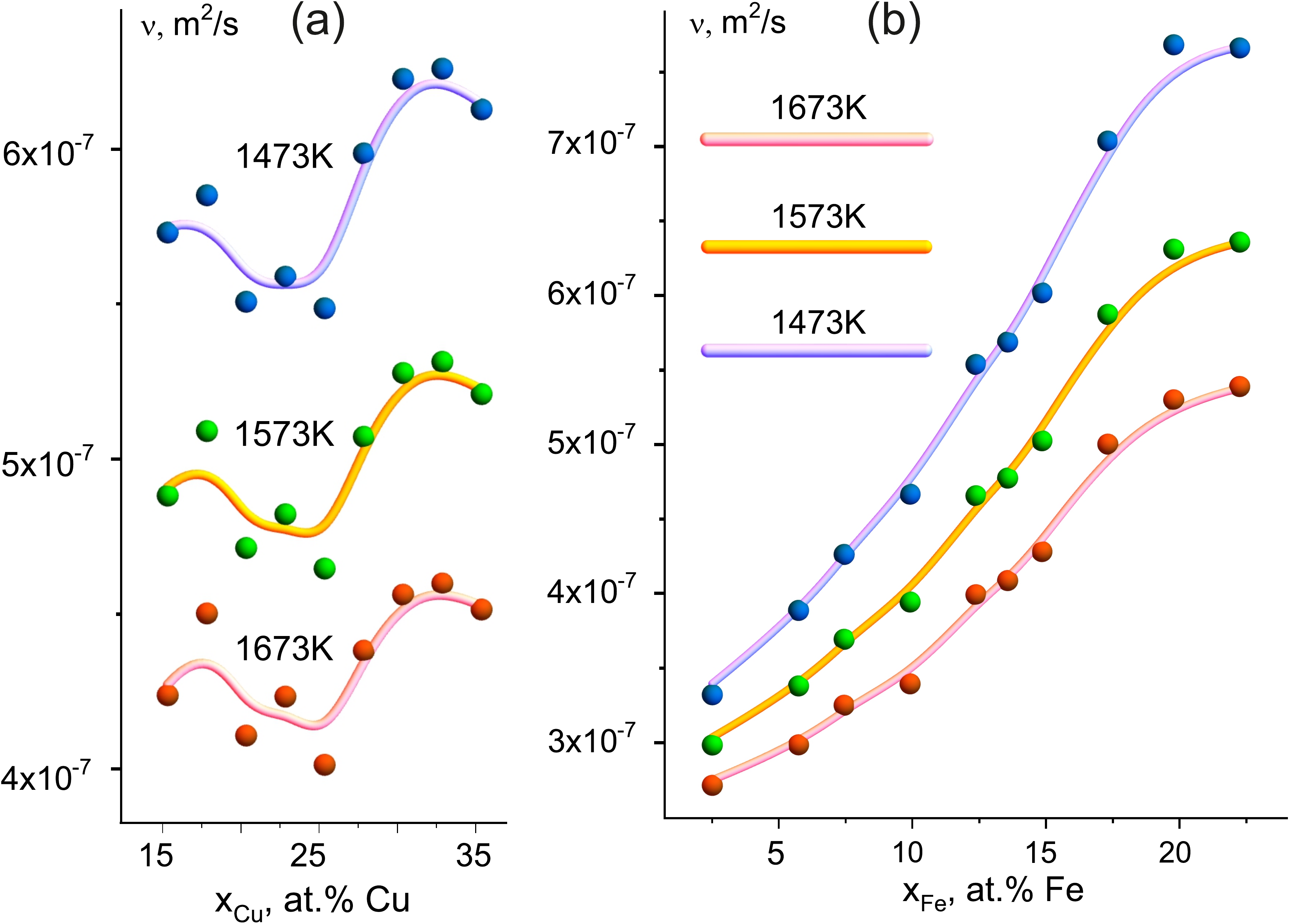}
  \caption{Concentration dependencies of the viscosity of Al-Cu-Fe melts at $x_{\rm Fe} = 12.5$ (a) and $x_{\rm Cu} = 25.5$ (b) and different temperatures.}\label{fig1}
\end{figure}
\begin{figure}[pos=t]
  \centering
  \includegraphics[width=\columnwidth]{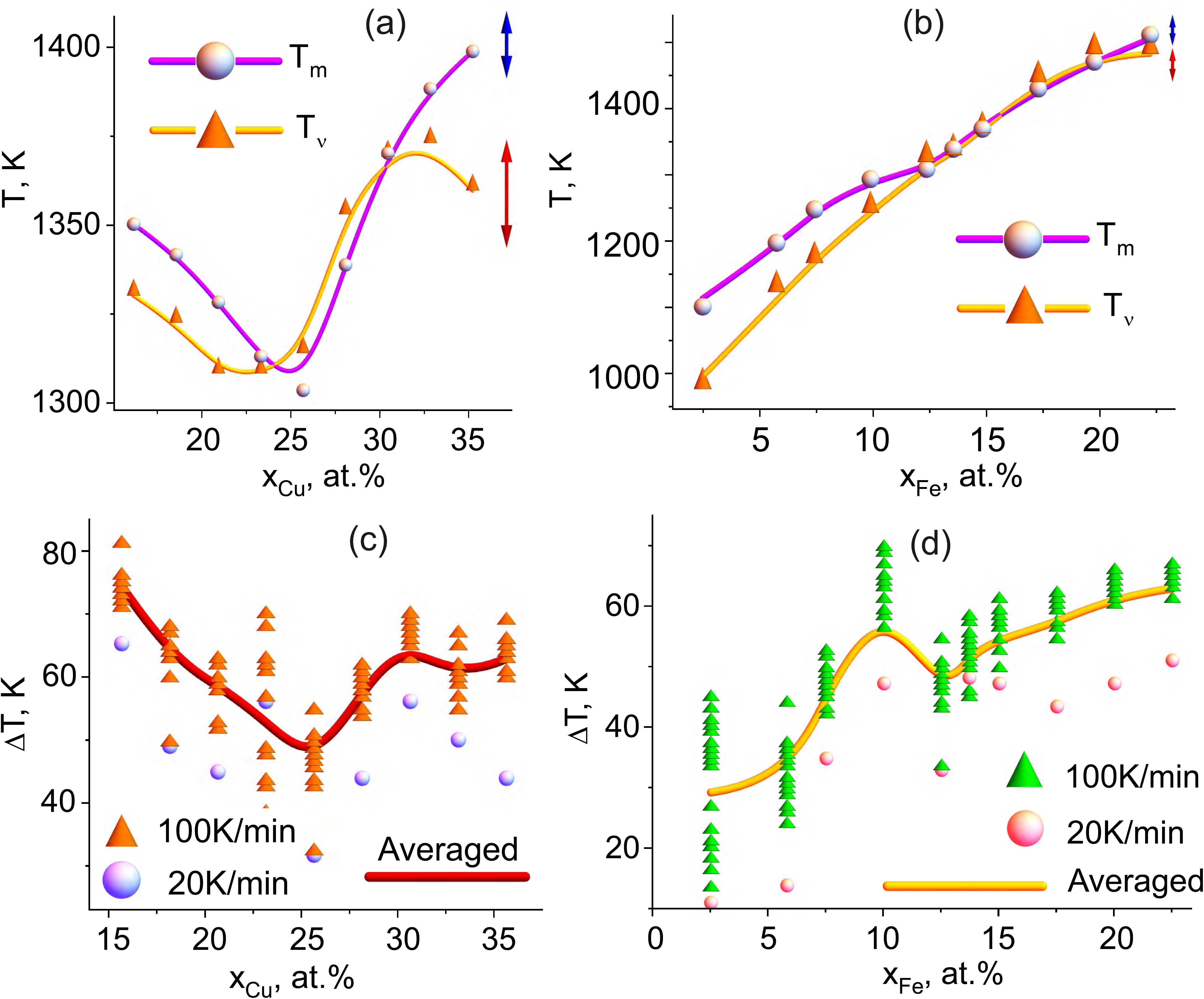}
  \caption{(a,b) Concentration dependencies of melting points $T_m$ (spheres) and lines of equal viscosity $T_{\nu}$  (cones); (c,d) Concentration dependencies of undercoolability obtained by cooling the melts from different temperatures at cooling rates of 100 K/min (cones) and 20 K/min (spheres). }\label{fig2}
\end{figure}
\begin{figure}[pos=t]
  \centering
  \includegraphics[width=\columnwidth]{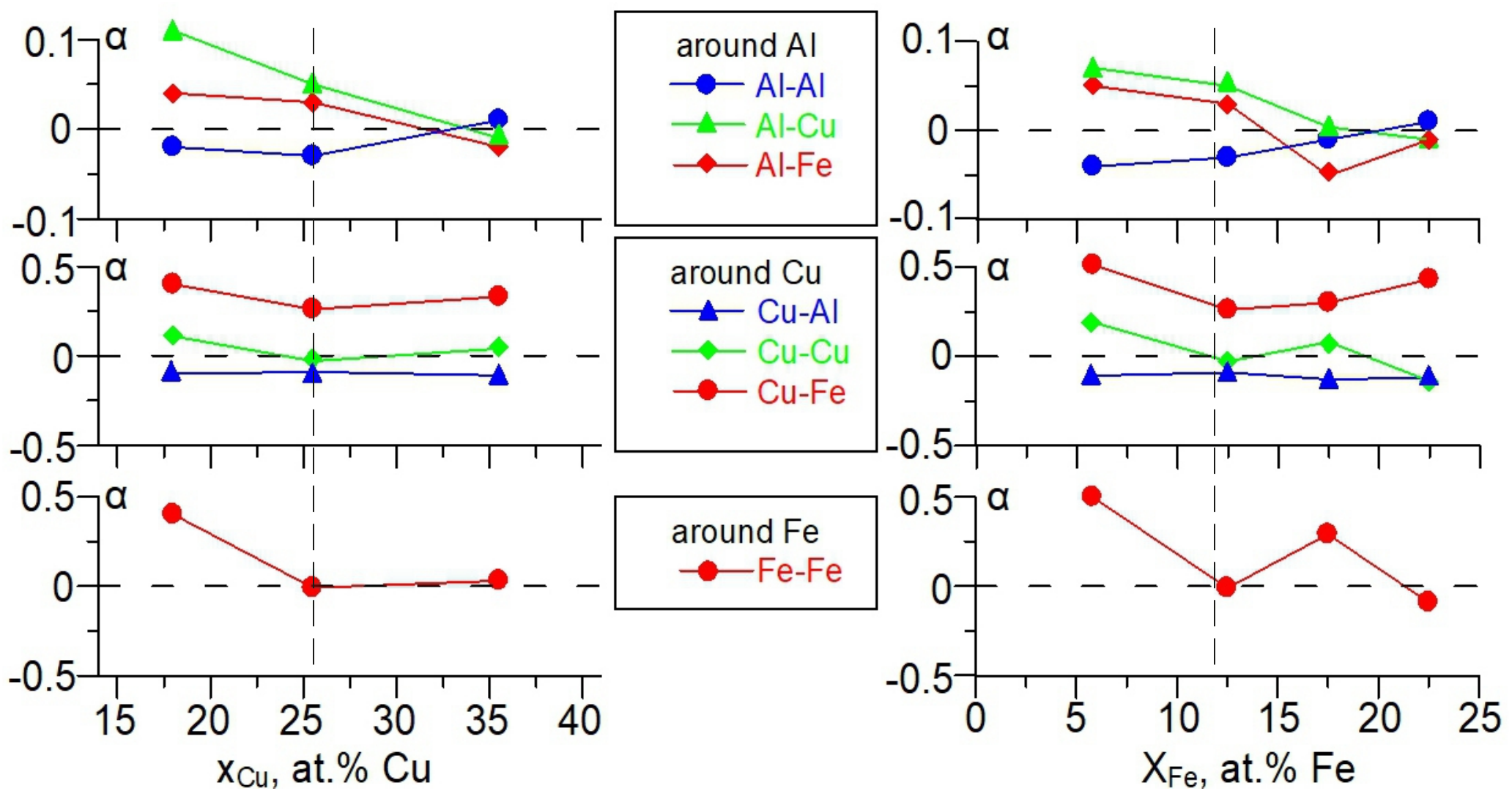}
  \caption{The concentration behavior of Warren-Cowley SRO parameters for Al-Cu-Fe melts extracted from \textit{ab initio} data. }\label{fig4}
\end{figure}

The experimentally obtained viscosity isotherms along two concentration cross-sections at $x_{\rm Fe} = 12.5$ at.\% and $x_{\rm Cu} = 25.5$ at.\% are shown in Fig.~\ref{fig1}. It can be seen from the figures that the replacement of aluminum by copper has little effect on the viscosity of Al-Cu-Fe melts in the investigated concentration range. On the concentration dependence $ \nu (x_{\rm Cu})$ obtained at $x_{\rm Fe} = 12.5$ at.\%, we observe weakly pronounced minimum at $x_{\rm Cu} = 25$ at.\% (Fig.~\ref{fig1}a); at $x_{\rm Cu} > 30$ at.\%,  viscosity is constant within the range of measurement error (Fig.~\ref{fig1}a). Increase of the iron content at a constant copper concentration leads to a sharp, but monotonous increase of viscosity (Fig.~\ref{fig1}b). An increase of the temperature leads to a weakening of the concentration dependence of the viscosity. However, the main features of $ \nu (x)$ dependence remain the same up to 1673~K (Fig.~\ref{fig1}a, b).

Fig.~\ref{fig2} shows lines of equal viscosity (red diamonds) obtained from $\nu (T)$ and liquidus lines (melting points $T_m$ versus concentration) determined from DTA curves during heating (black circles) (Fig.~\ref{fig2}a,b), as well as concentration dependencies of undercoolability obtained under cooling of  the melt from different temperatures at different cooling rates (Fig.~\ref{fig2}a, b). A line of equal viscosity is a concentration dependence $T_{\nu}(x)$ of temperatures at which the viscosity of a melts reaches a certain value. In our case, the viscosity of the i-phase near the melting point ($7.5 \cdot 10^{- 7 }m^{2}/s $) is chosen as such characteristic value. The lines of equal viscosity $T_{\nu}(x)$ and liquidus lines $T_m(x)$ demonstrate similar behaviour. At that, $T_{\nu, m} (x_{\rm Cu})$ develop minima near 25 at. \% Cu (Fig.~\ref{fig2}a), but $T_{\nu, m} (x_{\rm Fe})$ reveal sharp linear increase with increasing iron content $x_{\rm Fe}$ (Fig.~\ref{fig2}b).


It can be seen from the Fig.~\ref{fig2} that $T_{\nu=\nu_{\rm ico}}(x)$ (the temperature at which the viscosity of the melts studied is equal to the viscosity of the ico-phase near the liquidus temperature) coincides well with the liquidus line for each alloy. Noticeable discrepancy between these temperatures is observed for the alloys at $(x_{\rm Cu} > 30, x_{\rm Fe} = 12.5)$  and for $(x_{\rm Fe} < 12, x_{\rm Cu} = 25.5)$. For these alloys, the kinematic viscosity at the melting point is lower than $7.5\cdot 10^{-7}$ $m^2/s$; this viscosity value is achieved only under supercooling of the melts. Concentration dependencies of undercoolability $ \Delta T (x)$ obtained by cooling the melts from different temperatures at different cooling rates are shown in Fig.~\ref{fig2}c, d. Crystallization of the studied alloys proceeds at low undercoolings  $ \Delta T = 20 \div 80$ K.

\begin{figure}
  \centering
  \includegraphics[width=\columnwidth]{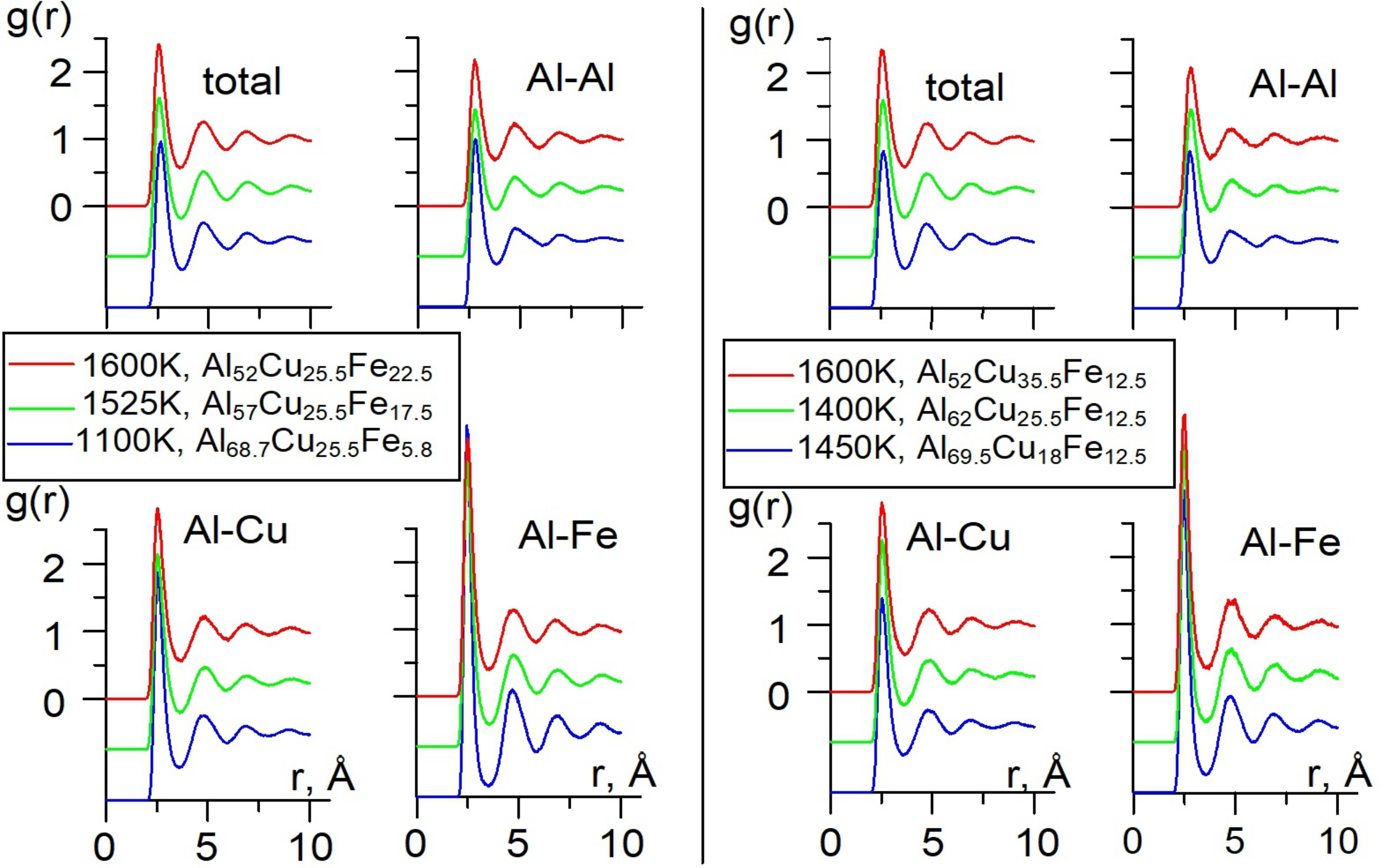}
  \caption{Total and partial RDF around Al in several Al-Cu-Fe melts at temperatures 100 K above corresponding melting points.}\label{fig3}
\end{figure}
\begin{figure}
  \centering
  \includegraphics[width=\columnwidth]{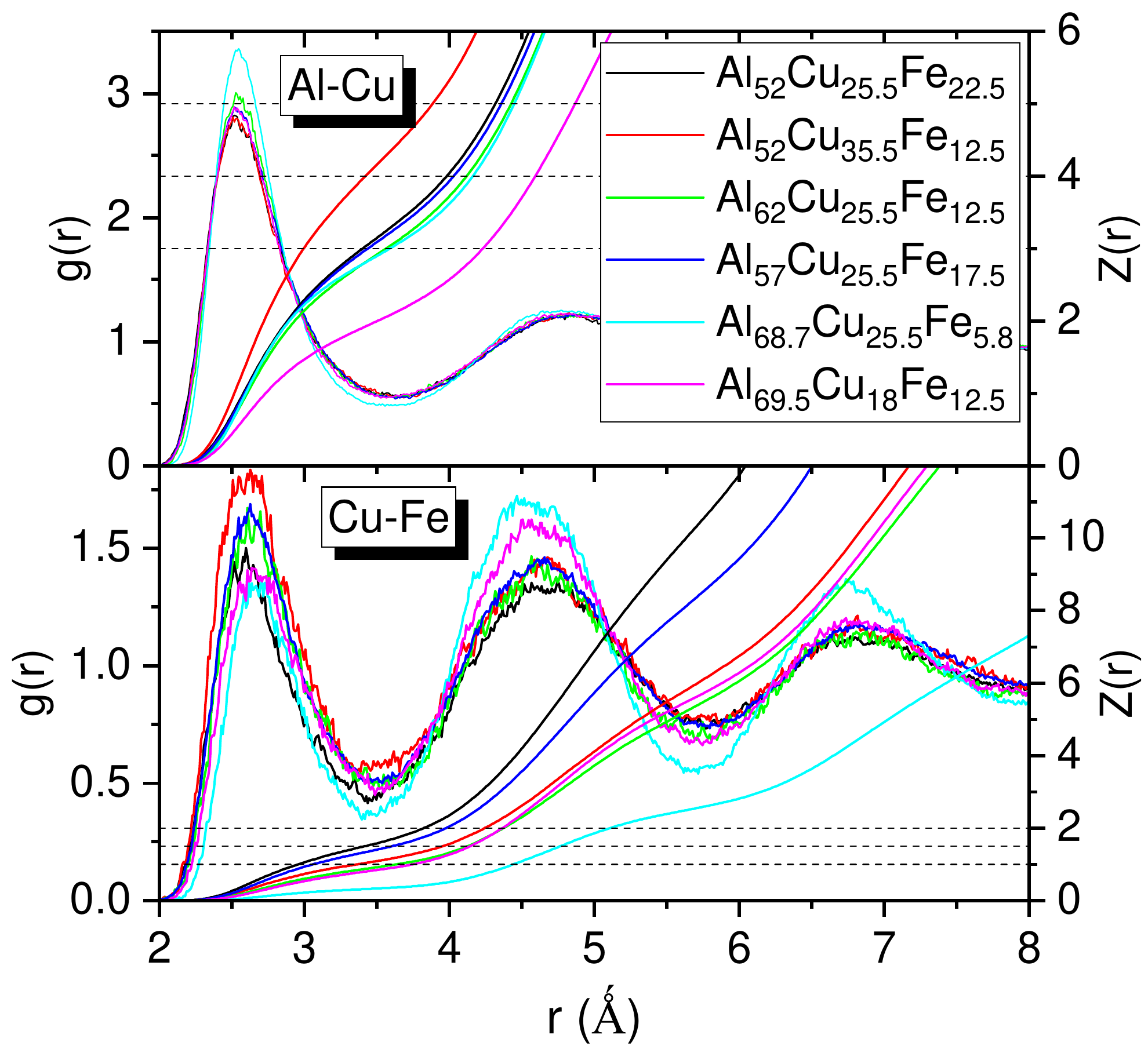}
  \caption{Detailed comparison of partial RDFs and partial coordinate dependent coordination numbers at different concentrations.}\label{figN}
\end{figure}

At $x_{\rm Fe} = 12.5$,  $ \Delta T (x_{\rm Cu})$  develops a minimum at $x_{\rm Cu} = 25$ that agrees well with both $T_{\nu}(x)$ and $T_{m}(x)$ behaviour (Fig.~\ref{fig2}a, c). At $x_{\rm Cu} = 25.5$,  $ \Delta T$ increases with $x_{\rm Fe}$; however, there is a kink at $ x_{\rm Fe} \approx 12.5 $ at. that is in agreement with the behaviour of $ \Delta T_m (x_{\rm Fe})$ (Fig.~\ref{fig2}b, d). Note that undercoolability of the alloys studied does not depend essentially on the initial temperature of the melts.   The cooling rate affects $\Delta T $ much more essentially, but in spite of that $\Delta T (x)$ are qualitatively the same at all cooling rates (Fig.~\ref{fig2}c, d).

Thus, near the melting point, viscosity of most alloys investigated has approximately the same value $7.5\cdot 10^{-7}$ $m^2/s$. Moreover, concentration dependencies $\nu (x)$ (Fig.~\ref{fig1}), $T_{\nu}(x)$  and  $T_{m}(x)$ (Fig.~\ref{fig2}) are qualitatively the same for all the studied compositions except $(x_{\rm Cu} > 30, x_{\rm Fe} = 12.5)$  and $(x_{\rm Fe} < 12, x_{\rm Cu} = 25.5)$. The concentration dependencies of undercoolability are also similar to the corresponding liquidus lines. Since both the viscosity and the undercoolability are structural-sensitive properties \cite{Sterkhova2014241,Beltyukov20151,Sterkhova2014250,Kamaeva2012InorgMater}, the features of their concentration dependencies may indicate structural changes in the melt. To analyze the structure of Al-Cu-Fe melts and its concentration changes, we perform AIMD simulations of six alloys: Al$_{68.7}$Cu$_{25.5}$Fe$_{5.8}$, Al$_{69.5}$Cu$_{18.0}$Fe$_{12.5}$, Al$_{62}$Cu$_{25.5}$Fe$_{12.5}$, Al$_{52}$Cu$_{35.5}$Fe$_{12.5}$, Al$_{57.0}$Cu$_{25.5}$Fe$_{17.5}$, and Al$_{52}$Cu$_{25.5}$Fe$_{22.5}$ at temperatures 100 K above corresponding $T_m$.

We place Fig.~\ref{fig4} (with Warren-Cowley SRO parameters $\alpha_{i-j}$ obtained from quantum molecular dynamics) near Figs.~\ref{fig1}-\ref{fig2} for comparing experiment and theory. The concentration dependencies of $\alpha_{Cu-Cu}$ and $\alpha_{Cu-Fe}$ in Fig.~\ref{fig4}  have local minimum (maximum) where there are features at experimental curves in Figs.~\ref{fig1}-\ref{fig2}. Moreover, $\alpha_{Al-Cu}$ and $\alpha_{Al-Fe}$ change their signs at the same compositions.  Accuracy limitations of quantum molecular dynamics do not allow drawing similar conclusions with respect to the other Warren-Cowley parameters. Below we discuss in more detail the Warren-Cowley SRO parameters for Al-Cu-Fe melts and how they were extracted from \textit{ab initio} calculations.

Using AIMD simulations, we calculate the total and partial radial distribution functions (RDF). In Fig.~\ref{fig3} we show total RDFs as well as partial RDFs around Al for all simulated alloys. We see that all total  RDFs are similar to that for simple liquids like Lennard-Jones one. We do not observe any special features of RDF like splitting of the peaks which allow predicting icosahedral phase formation~\cite{Ryltsev2015SoftMatt,Ryltsev2017SoftMatt,Englel2015Nature}.  To illustrate the dependence of RDFs from concentrations we draw them on the same graph, see Fig.~\ref{figN}, and plot the partial coordinate dependent coordination numbers. As seen from the Fig.~\ref{figN}, the clear difference between partial RDFs for Cu-Al and Cu-Fe takes place. The former is Lennard-Jones like but the latter demonstrates features, which are not typical for simple liquids. Namely, the heights of the first and the second peaks are almost the same. That means Cu and Fe avoid being nearest neighbours (see also the Warren-Cowley analysis below).

By using RDFs, one can determine useful structural characteristics, such as distance between nearest neighbors $r_{i-j}$  and coordination number $Z_{i-j}$. The former is determined from the positions of the first peaks of partial RDFs, and the latter is calculated by integrating the partial RDFs to their first minima; the area under the first peak of total RDF is equal to the number of nearest neighbors in the first coordination sphere. The values of these characteristics for simulated alloys are presented in Table~\ref{tab:table2}.

We remind that the number of atoms $$dZ(r)=\frac NV4\pi r^2 g(r) dr,$$ at a distance between $r$ and $r + dr$ from a given atom, where $N$ is the total number of atoms, $V$ is the volume and $g(r)$ is the total RDF. The partial RDF between the chemical species $i, j$ may be computed:
\begin{gather}
  g_{i-j}(r)=\frac{dZ_{i-j}(r)}{4\pi r^2\rho_{i}dr},\qquad \rho_{i}=\frac {N c_i} V,
\end{gather}
where $c_i$ is the concentration of atomic species $i$ and $Z_{i-j}(r)$ is the partial coordinate dependent coordination number.


\begin{table*}[pos=h]
\caption{Distances between nearest neighbors (r$ {} _ {i} $$ {} _ {-} $$ {} _ {j} $) and coordination numbers (Z${}_ {i} $${}_{-} $${}_ {j} $) in Al-Cu-Fe melts at temperatures 100 K above T${}_ {m} $${}_ {} $}. \label{tab:table2}
\begin{tabular}{|c|c|c|c|c|c|c|}
\hline
r, {\AA} & Al${}_{6}$${}_{9}$${}_{.}$${}_{5}$\newline Cu${}_{18}$\newline Fe${}_{12.5}$ \newline & Al${}_{62}$\newline Cu${}_{25.5}$\newline Fe${}_{12.5}$\newline & Al${}_{52}$\newline Cu${}_{35.5}$\newline Fe${}_{12.5}$\newline & Al${}_{68.7}$\newline Cu${}_{25.5}$\newline Fe${}_{5.8}$\newline & Al${}_{57}$\newline Cu${}_{25.5}$\newline Fe${}_{17.5}$\newline & Al${}_{52}$\newline Cu${}_{25.5}$\newline Fe${}_{22.5}$\newline \\ \hline
r(Al-Al) & 2.77 & 2.85 & 2.84 & 2.81 & 2.80 & 2.76 \\ \hline
r(Al-Cu) & 2.54 & 2.55 & 2.53 & 2.55 & 2.52 & 2.52 \\ \hline
r(Al-Fe) & 2.46 & 2.49 & 2.49 & 2.47 & 2.47 & 2.49 \\ \hline
r(Cu-Cu) & 2.50 & 2.55 & 2.54 & 2.56 & 2.53 & 2.54 \\ \hline
r(Cu-Fe) & 2.65 & 2.65 & 2.64 & 2.65 & 2.63 & 2.60 \\ \hline
r(Fe-Fe) & 2.98 & 2.28 & 2.23 & 2.90 & 2.63 & 2.34 \\ \hline
r(Al-tot) & 2.73 & 2.64 & 2.64 & 2.72 & 2.66 & 2.64 \\ \hline
r(Cu-tot) & 2.54 & 2.54 & 2.54 & 2.55 & 2.55 & 2.54 \\ \hline
r(Fe-tot) & 2.47 & 2.47 & 2.48 & 2.48 & 2.49 & 2.47 \\ \hline
r(tot) & 2.71 & 2.61 & 2.55 & 2.66 & 2.58 & 2.55 \\ \hline
Z &  &  &  &  &  &  \\ \hline
Z(Al-Al) & 8.9 & 7.9 & 6.5 & 9.1 & 7.5 & 6.8 \\ \hline
Z(Al-Cu) & 2.0 & 3 & 4.5 & 3.0 & 3.3 & 3.4 \\ \hline
Z(Al-Fe) & 1.5 & 1.5 & 1.6 & 0.7 & 2.4 & 3 \\ \hline
Z(Cu-Al) & 8.1 & 7.3 & 6.9 & 8.1 & 7.4 & 6.8 \\ \hline
Z(Cu-Cu) & 1.7 & 2.5 & 4.1 & 2.2 & 2.7 & 3.4 \\ \hline
Z(Cu-Fe) & 0.8 & 1.0 & 1.0 & 0.3 & 1.4 & 1.5 \\ \hline
Z(Fe-Fe) & 0.8 & 1.4 & 1.4 & 0.3 & 1.4 & 2.8 \\ \hline
Z(Al-tot) & 12.5 & 12.4 & 12.6 & 12.7 & 13.0 & 13.2 \\ \hline
Z(Cu-tot) & 10.6 & 10.8 & 12.0 & 10.6 & 11.5 & 11.7 \\ \hline
Z(Fe-tot) & 10.7 & 11.1 & 11.6 & 10.3 & 11.3 & 11.4 \\ \hline
Z(tot) & 11.9 & 11.8 & 12.3 & 12.1 & 12.3 & 12.4 \\ \hline
\end{tabular}
\end{table*}

We see from the Table~\ref{tab:table1} that replacing Al atoms by either Cu or Fe ones leads to a decrease in the radius of the first coordination sphere, while the coordination number is almost constant and equal to that for closed packed structures ($Z = 12$). Both $r_{\rm Al - Al}$ and $r_{\rm Cu - Cu}$ nearest neighbors distances are close to the radii of the first coordination spheres in liquid Al~\cite{brillo20064008} and Cu~\cite{cahn1982structure}, respectively; for  $r_{\rm Fe - Fe}$ and the Fe melt, such relation does not take place. It should be noted that simulation data for partial RDFs with iron atoms are the least accurate ones since the concentration of iron  is rather low in most of the alloys studied. The most pronounced concentration variations are observed for the radius of the first coordination shell around Al atoms, which are primarily related to the concentration dependence of the distance between Al atoms.

In order to analyse chemical SRO, we calculate Warren-Cowley parameters~\cite{Warren1951JApplPhys,wang2017115}, which are extracted from both total and partial coordination numbers as:
$$
\alpha_{i-j}=1-\frac{Z_{i-j}}{Z_{i-\mathrm{total}}x_{j}}.
$$
Here $Z_{i-\mathrm{total}}$ and $Z_{i-j}$ are respectively the total and partial coordination numbers, $x_{j}$ is the $j$ atom concentration. For a random atom distribution, $\alpha$$_{i-j}$ are equal to zero. The negative $\alpha$$_{i-j}$ means effective attraction between $i$ and $j$ species, while the positive $\alpha$$_{i-j}$ reflects repulsion. Calculated $\alpha$$_{i-j}$ values for Al-Cu-Fe melts are presented in Tab.~\ref{tab:table3}; Fig.~\ref{fig4} demonstrates their concentration behaviour.

\begin{table*}[pos=h]
\caption{Warren-Cowley SRO parameters in Al-Cu-Fe melts at temperatures 100~K above~T$_ {m} $}\label{tab:table3}
\begin{tabular}{|c|c|c|c|c|c|c|} \hline
$\alpha$(i-j) & Al${}_{69.5}$\newline Cu${}_{18}$\newline Fe${}_{12.5}$ & Al${}_{62}$\newline Cu${}_{25.5}$\newline Fe${}_{12.5}$ & Al${}_{52}$\newline Cu${}_{35.5}$\newline Fe${}_{12.5}$ & Al${}_{68.7}$\newline Cu${}_{25.5}$\newline Fe${}_{5.8}$ & Al${}_{57}$\newline Cu${}_{25.5}$\newline Fe${}_{17.5}$ & Al${}_{52}$\newline Cu${}_{25.5}$\newline Fe${}_{22.5}$ \\ \hline
$\alpha$(Al-Al) & -0.02 & -0.03 & 0.008 & -0.04 & -0.01 & 0.009 \\ \hline
$\alpha$ (Al-Cu) & 0.11 & 0.05 & -0.006 & 0.07 & 0.004 & -0.01 \\ \hline
$\alpha$ (Al-Fe) & 0.04 & 0.03 & -0.02 & 0.05 & -0.05 & -0.01 \\ \hline
$\alpha$ (Cu-Al) & -0.099 & -0.09 & -0.11 & -0.11 & -0.13 & -0.12 \\ \hline
$\alpha$ (Cu-Cu) & 0.11 & -0.03 & 0.04 & 0.19 & 0.08 & -0.14 \\ \hline
$\alpha$ (Cu-Fe) & 0.40 & 0.26 & 0.33 & 0.51 & 0.30 & 0.43 \\ \hline
$\alpha$ (Fe-Fe) & 0.40 & -0.009 & 0.03 & 0.5 & 0.29 & -0.09 \\ \hline
\end{tabular}
\end{table*}

\begin{figure*}
  \centering
  \includegraphics[width=\textwidth]{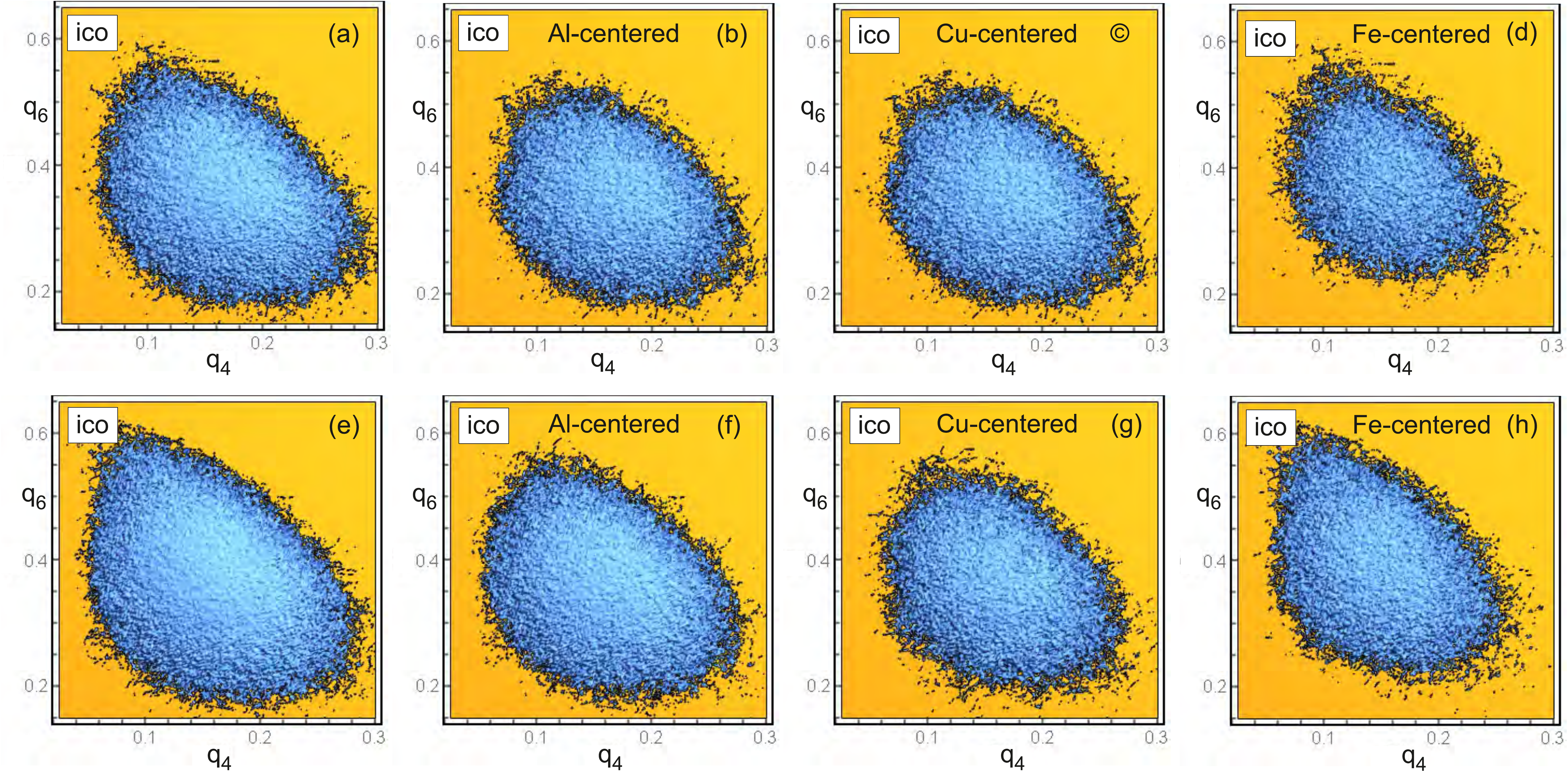}
  \caption{Local orientational order of the ${\rm Al_{52}Cu_{25.5}Fe_{22.5}}$ alloy on the $q_4\--q_6$ plane. Bond orientational order parameters (BOOPs) were calculated via 12 nearest neighbours for Al, Cu and Fe-centered atoms. Points on the pictures correspond to the $q_4-q_6$ values for each atom;  $(q_4,q_6)$-point for the perfect icosahedron is indicated as ico. Temperature of the system is $T=1600$~K for (a)-(d) and 1000~K for (e)-(h). The system at $T=1000$ (deep undercooling condition) was cooled from 1600~K  for 10000~fs and then relaxed at $T=1000$ for 5000~fs. This is not enough for equilibration under such undercooling but the tendency is already seen --  local icosahedral order becomes much more pronounced than at 1600K.}\label{fig:PQ}
\end{figure*}

As seen from Tab.~\ref{tab:table3} and Fig.~\ref{fig4}, the strongest chemical interaction in Al-Cu-Fe melts is the repulsion between Cu and Fe atoms that is in agreement with the results obtained for these alloys in solid state~\cite{brand2000662,brand2001210}. An increase of $x_{\rm Cu}$ at $x_{\rm Fe} = 12.5$ as well as an increase of  $x_{\rm Fe}$ at $x_{\rm Cu} = 25.5$ lead to a change of effective interatomic interaction for Al:  the Al-Al attraction, Al-Cu repulsion and Al-Fe repulsion decrease with $x$ and change their signs in the vicinity of i-phase stoichiometry (this is especially pronounced for Al-Fe interaction). At high concentrations of either copper or iron atoms around Al ones, a nearly random distribution of atoms without pronounced chemical ordering is observed. Concentration changes of chemical SRO around Cu atoms show that the weak attraction between Al-Cu atoms does not depend on the concentration of Cu and Fe for all the melts studied. The pronounced repulsion between Cu and Fe atoms has minimal values at i-phase stoichiometry; at this concentration, a random distribution of both Al-Cu and Cu-Cu bonds is observed. At high Al concentrations, Fe atoms avoid each other in Al-Cu-Fe melts.

\begin{figure}
  \centering
  \includegraphics[width=\columnwidth]{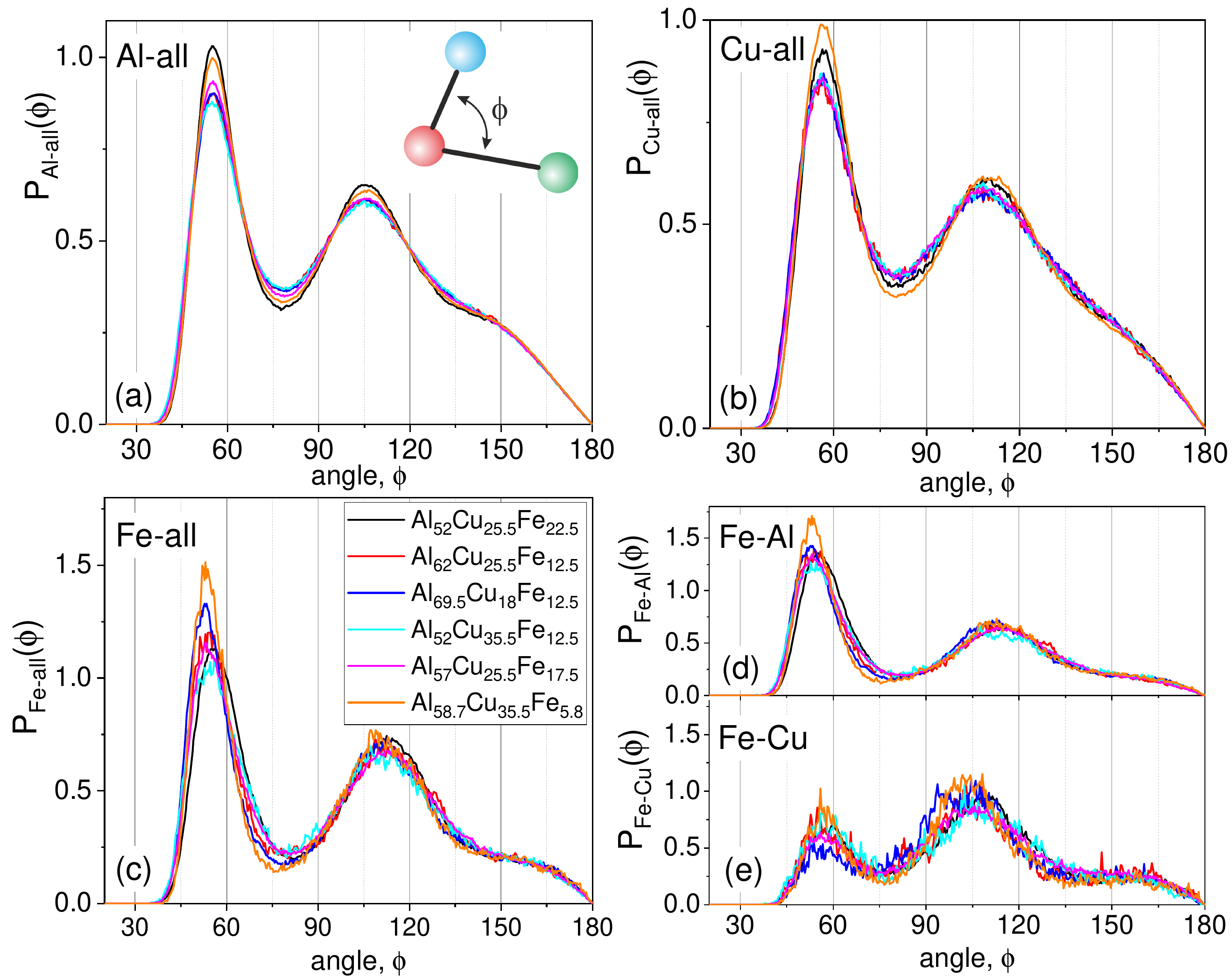}
  \caption{Bond-angle distribution function for Al-Cu-Fe melts extracted from AIMD data.}\label{fig5}
\end{figure}

Important structural information beyond pair distribution can be extracted from  bond-angle distribution function (BADF), which defines probability distribution $P(\theta)$ for the angle $\theta$ between a chosen particle and its two nearest neighbours. The BADF for Al-Cu-Fe alloys, extracted from AIMD data, are shown in Fig.~\ref{fig5}. We see that distributions around both Al and Cu atoms are similar to those for other closed packed systems like metallic melts of Lennard-Jones liquid \cite{Ryltsev2013PRE,Klumov2018JCP}: there is the main peak at $\theta \approx 60$ and the second less pronounced one at $\theta \approx 110$. However, the situation for Fe-centered distributions is quite different. We see that the first BADF peak is much more pronounced so that the $P(\theta)$ at the first minimum is close to zero. Analysis of partial BADF reveals that this effect is mainly due to Al-Fe-Al angles (Fig.~\ref{fig5}d). That indicates strong chemical interaction between Al and Fe. Whereas, BADFs for Cu-Fe-Cu angles demonstrate that Cu and Fe avoid forming triangles with each other (the peak at $\theta \approx 60$ is relatively weak) that agrees with the results obtained from Warren-Cowley parameters (see Tab.~\ref{tab:table3}). Note that BADFs for all studied melts demonstrate weak concentration dependence.

\section{Discussion}

As suggested in~\cite{brand2000662,brand2001210}, SRO in solid Al-Cu-Fe alloys is characterized by strong interaction between Al-Cu and Al-Fe pairs, while Fe atoms avoid being in the nearest environment with Cu atoms. Such interaction leads to the formation of i-phase whose structure consists of Bergman clusters~\cite{Tsai2008}. Such a cluster for Al-Cu-Fe i-phase includes 33 atoms: Cu atom in the center is surrounded by an icosahedron built of Al atoms; the icosahedron, in turn, is surrounded by a dodecahedron built of Cu, Al and Fe atoms.


To study the possible correlation between the structure of i-phase and SRO of its melt, we perform Voronoi analysis~\cite{voropp,vorotop,Stukowski2012ovito} of AIMD configurations. We find that topologically ideal icosahedra (polyhedra with Voronoi index $\langle0,0,12,0\rangle$) are almost absent in the structure of all the melts studied. However SRO in Al-Cu-Fe melts is polytetrahedral, that is, it is presented by Kasper polyhedra, which can be treated as icosahedra with topological defects -- disclinations~\cite{Cheng2011ProgMateSci}. For all compositions, the most favorable polyhedron (about 3 \%) is $\langle0,3,6,4\rangle$, which is strongly distorted Kasper polyhedron Z13.  Thus, we argue that polytetrahedral clusters strongly distorted by thermal fluctuations are the main structural motifs in Al-Cu-Fe melts. Note that Kasper polyhedra are also suggested to be  structural motifs in glass-forming metallic liquids and glasses \cite{Ryltsev2018JCP, Klumov2018JCP, Sheng2006Nature,Wang2016JPhysChemB,Guerdane2008PRL}. That means polytetrahedral order is probably inherent for metallic alloys.

The results of Voronoi analysis suggest that there is no clear relation between the local structure in a solid phase and topology of Voronoi polyhedra in the corresponding melt. To validate this results, we study local orientational order in the melts by using bond orientational order parameters (BOOPs), which are widely used in structural analysis of condensed matter systems \cite{Ryltsev2016JCP,Klumov2018JCP,Fomin2014,Hirata2013Science,Klumov2016JETPLett}. Detailed description of the method can be found in \cite{Steinhardt1983PRB}. Briefly, we calculate the rotational invariants $q_l$, of rank $l$ for each atom using the fixed coordination number (${\rm CN}=12$). These invariants are uniquely determined for any polyhedron including the elements of any crystalline structure. Here we use $q_4$, $q_6$ as the the most informative ones; their values for a number of close-packed structures are presented in Table~\ref{tableQ}.
\begin{table}
\centering
\caption{Rotational invariants (RI) $q_l$ of a few perfect clusters calculated via fixed number of nearest neighbors (NN): hexagonal close-packed (hcp), face centered cubic (fcc), icosahedron (ico). Additionally, mean RI for the Lennard-Jones melt are shown for the comparison.}
\begin{tabular}{|c|c|c|}
\hline 
cluster  & \quad $q_{4}$ & \quad $q_{6}$  \\ \hline
hcp (12 NN) & 0.097 & 0.485  \\ \hline
fcc (12 NN) & 0.19  & 0.575  \\ \hline
ico (12 NN) & $1.4 \times 10^{-4}$ & 0.663  \\ \hline
LJ melt (12 NN) & $\approx $0.155  & $\approx $0.37   \\ \hline
\end{tabular}
\label{tableQ}
\end{table}

 In Fig.~\ref{fig:PQ}(a-d) we show BOOP probability distribution for ${\rm Al_{52}Cu_{25.5}Fe_{22.5}}$ alloy on the $q_4\--q_6$ plane at $T=1600$ K. Each point on the pictures correspond to $(q_4, q_6)$ value for certain atom. Comparing the distributions with $(q_4\,q_6)$ value for perfect icosahedron (labeled as ico), we conclude that BOOPs do not demonstrate pronounced icosahedral ordering; only some traces of such ordering are observed for Fe-centered local clusters.


Thus, we observe that the melt of locally icosahedral quasicrystal phase does not contain topological icosahedra due to strong thermal fluctuations, which smear fine structural properties of local polyhedra. It can be expected that a noticeable amount of icosahedral clusters in Al-Cu-Fe liquid will occur in the supercooled state where local order is much more pronounced. However, investigation of deeply supercooled liquids by AIMD is an extremely difficult task due to much time needed to form an equilibrated structure \cite{Zhang2016JCP}. This problem can be overcome by utilizing semi-empirical potentials fitted to the \textit{ab initio} and/or experimental data. Development of such potential for Al-Cu-Fe system is a matter of the forthcoming paper. However, in this work we perform preliminary investigation of supercooled ${\rm Al_{52}Cu_{25.5}Fe_{22.5}}$ liquid by using AIMD simulations.  A high-temperature configuration of the system at $T=1600$~K was cooled down to 10,000~fs for and then relaxed at this temperature for 5,000~fs. Of course, this is not enough for equilibration of the system under such undercooling but the tendency of structural relaxation can be observed. Indeed, in Fig.~\ref{fig:PQ}(a-d) we show BOOP distributions for ${\rm Al_{52}Cu_{25.5}Fe_{22.5}}$ aloy at $T=1000$~K. It is seen from the picture that  deeply supercooled alloy, even being in a strongly underelaxed state, demonstrates much more pronounced icosahedral ordering in comparison with the equilibrium melt.

As we have shown above, orientational order of the nearest neighbors in the solid state is essentially destroyed in the melt. However, some correlation between the structure of solid and liquid phases can be extracted from more rough structural characteristics, such as RDF and BADF, which are more stable against thermal fluctuations. Indeed, analysis  of chemical SRO performed with both RDF and BADF reveals that the main features of interatomic interactions in solid Al-Cu-Fe alloys are the same as for their melts. First, there is pronounced repulsion between Cu and Fe atoms, which is practically concentration-independent (see Tab.~\ref{tab:table3}). Second, strong chemical interaction between Fe and Al takes place (see Fig.~\ref{fig5}d). Both these properties are characteristic for solid Al-Cu-Fe alloys~\cite{brand2000662,brand2001210}. Other features of the chemical interaction in ternary Al-Cu-Fe melts are closely related to  concentration changes of the structure in binary Al-Cu and Al-Fe systems. The structure of binary  Al-Cu and Al-Fe melts has been intensively studied by both X-ray diffraction method and molecular dynamics simulations in a wide range of compositions~\cite{brillo20064008,Kang2012PRL,jakse2016PRB,Jingyu1998}. In Al-Cu melts, an increase of copper concentration from 10 to 40 at.\% Cu leads to the change of SRO associated with the change of the type of interatomic interaction from  homo-coordinated (with preferred Al-Al and Cu-Cu bonds) to hetero-coordinated (with preferred Al-Cu bonds), while at $x_{\rm Cu}\approx 25$ at.\% there is a sharp decrease in the average distances between Al-Al and Cu-Cu pairs~\cite{jakse2016PRB}. In Al-Fe melts, at $x_{\rm Fe} > 25$ at.\% a shoulder on statics structure factor occurs that indicates a change in the type of structural ordering ~\cite{Jingyu1998}. These features of the structure of Al-Fe and Al-Cu melts can explain the changes of sign of the effective interaction between Al-Cu and Al-Fe pairs observed in the vicinity of i-phase stoichiometry. Note that, at this concentration, we observe the largest distance between the Al atoms in the first coordination shell.

Peculiarities of SRO discussed above allow understanding concentration variations of viscosity in Al-Cu-Fe melts. When describing the mechanisms of viscosity in metal melts, two factors are usually considered: kinetic factor and that associated with the breaking of interatomic bonds~\cite{jakse2016PRB}. Since the change in temperature does not affect essentially concentration behavior of the viscosity (Fig.~\ref{fig1}), we assume that it is mainly determined by the interatomic interaction in the system. The weakness of interaction between Cu and Fe leads to the fact that the concentration dependencies of the viscosity in Al-Cu-Fe system along two cross-sections considered are mainly determined by the mechanisms determining viscosity variations in binary Al-Cu and Al-Fe melts~\cite{Beltyukov20151,Weimin1552}. According to available literature data~\cite{Weimin1552}, the viscosity of Al-Cu melts does not change noticeably at $x_{\rm Cu} < 20$, but increase at $x_{\rm Cu} > 20$. In Al-Fe system, an increase of $x_{\rm Fe}$ leads to a sharp monotonic growth of the viscosity at $x_{\rm Fe}> 2$ at. \% Fe~\cite{Beltyukov20151}.

 However, we reveal some properties of the viscosity in the triple system, which are not observed in corresponding binary alloys. First, high viscosity values in Al-Cu-Fe melts near the liquidus (compared to those in binary systems) are observed for the compositions in the vicinity of i-phase stoichiometry (Fig.~\ref{fig2}). Second, concentration dependencies of Al-Cu-Fe viscosity develop minima at this composition (Fig.~\ref{fig1}), that is also not typical for binary Al-Cu and Al-Fe systems. These features of the viscosity concentration behavior are related with the concentration variation of  interatomic interaction. Indeed, at $x_{\rm Cu} = 25.5$ and $x_{\rm Fe} = 12.5$, chemical interaction is minimal in the melts. Local minima at this composition are also observed on concentration dependencies of the melting temperature. The data on undercoolabiliy obtained during solidification at different cooling rates 20-100 K/min reveal that the concentration variations of chemical interaction described above affect noticeably the initial stage of solidification (the temperature at which the formation of a solid phase begins in the system). We argue that obtained structural peculiarities of Al-Cu-Fe melts are important for fabricating Al-Cu-Fe i-phase from the melt.

\section{Conclusions}

We present a detailed study of the structure of Al-Cu-Fe melts and its structural-sensitive properties, such as viscosity and undercoolability.  We focus on two concentration cross-sections Al$_{57+x}$Cu$_{30.5-x}$Fe$_{12.5}$ and Al$_{52+x}$Cu$_{25.5}$Fe$_{22.5-x}$, ($x=0-20$ at.\%), which both contain the stoichiometry composition of icosahedral quasicrystal phase.

We observe that, along the cross-sections studied, the concentration dependencies of the viscosity are close to those in binary Al-Cu and Al-Fe melts. The maximal value of the viscosity is observed in the vicinity of i-phase stoichiometry. We experimentally observe that concentration dependencies of the viscosity, concentration lines of equal viscosity and concentration dependencies of the temperatures of the first stage of crystallization reproduce qualitatively the liquidus line. All of these characteristics develop minima at concentration corresponding to i-phase stoichiometry.

Using AIMD simulations we also study the structure of the Al-Cu-Fe melts at temperatures which are slightly above the liquidus line.  We show that SRO in Al-Cu-Fe melts is mainly presented by distorted polytetrahedral Kasper polyhedra with the number of nearest neighbors close to that in closed packed systems ($Z = 12$). However, topologically perfect icosahedra are almost absent an even i-phase stoichiometry that suggests the topological structure of local polyhedra does not survive upon melting. However, structural characteristics extracted from radial distribution function and bond-angle distribution function are more stable against thermal fluctuation and thus allow detecting structural heredity between solid and liquid phases. In particular, by analysing the Warren-Cowly Warren-Cowley parameters, we show that the main features of effective interatomic interaction in Al-Cu-Fe system are the same for both liquid and solid states. For example, there is a pronounced repulsion between Cu and Fe atoms, which is almost independent of the concentration of the components in the melt. Moreover, the  chemical SRO of the melts changes qualitatively in the vicinity of i-phase stoichiometry.

The results obtained demonstrate a relation between the structure of Al-Cu-Fe melt, its structural sensitive properties and the tendency to form icosahedral phase.

\section{Acknowledgments}
This work was supported by Russian Science Foundation (grant 18-12-00438). We thank  ``Uran'' supercomputer of IMM UB RAS for access.  Model simulations have been carried out using computing resources of the federal collective usage center Complex for Simulation and Data Processing for Mega-science Facilities at NRC ``Kurchatov Institute'', http://ckp.nrcki.ru/.

\bibliographystyle{model1-num-names}
\bibliography{our_bib}
\end{document}